\documentclass[sn-mathphys,Numbered]{sn-jnl}

\usepackage{graphicx}%
\usepackage{multirow}%
\usepackage{amsmath,amssymb,amsfonts}%
\usepackage{amsthm}%
\usepackage{mathrsfs}%
\usepackage[title]{appendix}%
\usepackage{xcolor}%
\usepackage{textcomp}%
\usepackage{manyfoot}%
\usepackage{algorithm}%
\usepackage{algorithmicx}%
\usepackage{algpseudocode}%
\usepackage{listings}%

\theoremstyle{thmstyleone}%
\theoremstyle{thmstyletwo}%

\theoremstyle{thmstylethree}%

 \usepackage{afterpage} 
 
\DeclareRobustCommand{\txr}[1]{{\textcolor{black}{#1}}} 

\begin{document}

\title[Article Title]{
Deterministic Random Walk Model in NetLogo and the Identification of Asymmetric Saturation Time in Random Graph
}


\author[1]{\fnm{Ayan} \sur{Chatterjee}}
\equalcont{These authors contributed equally to this work.}

\author[2]{\fnm{Qingtao} \sur{Cao}}
\equalcont{These authors contributed equally to this work.}

\author[3]{\fnm{Amirhossein} \sur{Sajadi}}

\author*[1,4]{\fnm{Babak} \sur{Ravandi}}\email{b.ravandi@northeastern.edu}


\affil[1]{\orgdiv{Network Science Institute}, \orgname{Northeastern University}, \orgaddress{\city{Boston}, \postcode{02215}, \state{MA}, \country{USA}}}

\affil[2]{\orgdiv{Department of Mechanical and Industrial Engineering}, \orgname{Northeastern University}, 
\orgaddress{\city{Boston}, \postcode{02215}, \state{MA}, \country{USA}}}

\affil[3]{\orgdiv{Renewable and Sustainable Energy Institute}, \orgname{University of Colorado Boulder}, \orgaddress{\city{Boulder}, \postcode{80303}, \state{CO}, \country{USA}}}

\affil[4]{\orgdiv{Department of Physics}, \orgname{Northeastern University}, \orgaddress{\city{Boston}, \postcode{02215}, \state{MA}, \country{USA}}}

\abstract{
Interactive programming environments are powerful tools for promoting innovative network thinking, teaching science of complexity, and exploring emergent phenomena. 
This paper reports on our recent development of the deterministic random walk model in NetLogo, a leading platform for computational thinking, eco-system thinking, and multi-agent cross-platform programming environment. 
The deterministic random walk is foundational to modeling dynamical processes on complex networks. Inspired by the temporal visualizations offered in NetLogo, we investigated the relationship between network topology and diffusion saturation time for the deterministic random walk model. 
Our analysis uncovers that in Erd\H{o}s-R\'{e}nyi graphs, the saturation time exhibits an asymmetric pattern with a considerable probability of occurrence. 
This behavior occurs when the hubs, defined as nodes with relatively higher number of connections, emerge in Erd\H{o}s-R\'{e}nyi graphs. 
Yet, our analysis yields that the hubs in  Barab\'{a}si-Albert model stabilize the the convergence time of the deterministic random walk model. 
These findings strongly suggest that depending on the dynamical process running on complex networks, complementing characteristics other than the degree need to be taken into account for considering a node as a hub. 
We have made our development open-source, available to the public at no cost at \hyperlink{https://github.com/bravandi/NetLogo-Dynamical-Processes}{https://github.com/bravandi/NetLogo-Dynamical-Processes}.
}


\maketitle

\section{Introduction}

Study of complex systems and the desire to understand and uncover complex phenomena dates back to 18th century when Sir Isaac Newton formulated the famous three-body problem \cite{newton1833philosophiae}. 
Ever since, this field has evolved itself into the realm of complex network science, as an emerging research field that explains the foundational necessities for describing and understanding highly heterogeneous and interconnected complex systems \cite{strogatz2001exploring}. 

Complex networks are often distinguished by exhibiting emergent phenomena that cannot be solely described by the sum of the behavior of their constituent subsystems/subnetworks \cite{complexity_guided_tour}. 
Extensive research effort within this realm over the past two decades has contributed to the development of the knowledge and the theories that shed light on the understanding of complex dynamical processes like the spread of epidemics \cite{keeling2005networks,newman2002spread,pastor2015epidemic}, the extent of controllability in networked systems \cite{liu2011controllability,heydari2019analysis}, synchronization in complex power networks, \cite{sajadi2022synchronization,motter2013spontaneous,dorfler2012synchronization}, and the pollination patterns in ecology \cite{olesen2007modularity,olesen2002geographic}, just to name a few. 
Nonetheless, the education of network thinking and applications of complex networks to a broad range of engineering and science disciplines has remained an open challenge because of the inherent sophistication associated with this subject and its foundational analytical complexity which many may find to be acute. In this paper, we address this topic by borrowing the interactive element and temporal nature of NetLogo \cite{tisue2004netlogo} to promote network thinking in both teaching and research. 
We believe enhancing the ability to communicate and model complexity is an essence to develop the next generation of interdisciplinary thinkers, necessary for tackling the grand challenges of our century, particularly that the unprecedented connectivity around us continues to escalate the complexity of all disciplines.

Amongst all the aspects of complex networks, random walk theory remains as one of the most important developments \cite{random_walk_maps,random_walk_complex_nets}. 
Random walk is a fundamental model for explaining the observable behavior of many stochastic processes, with its application spanning over various scientific fields from ecology and psychology to physics and chemistry \cite{burioni2005random,masuda2017random}. 

The saturation time on random walk is quantified by the number of steps from the initiation to the final convergence of the walk, and it has been mainly studied in the context of information spread over networks \cite{jia2010method,nmi11,Milli2018,Elmaghraby1997}. 
To the best of our knowledge, this is the first study on the relationship between network topology and saturation time of random walk processes.

The contributions of this work are threefold. 
First, we report on our recent contribution to NetLogo by developing the deterministic random walk model which is very practical for education of network science fundamentals. 
NetLogo is a leading platform for complex computational analytics, eco-system modeling, and multi-agent cross-platform programming \cite{tisue2004netlogo}. 
Second, our development uncovers that in Erd\H{o}s-R\'{e}nyi (ER) graphs \cite{erdos59a} the saturation time of the deterministic random walk exhibits an asymmetric pattern with a considerable probability of occurrence, highlighting the relationship between network structure and expected outcome of dynamical processes. 
Third, leveraging our development, we propose that the identification of hubs in networks may not always be based on having a large degree, depending on the dynamical processes running over a network. 
Identifying hubs only based on degree is commonly used to analyze complex networks. 
Yet, our analysis demonstrates that the inter-connectivity between the hubs influences the saturation time in deterministic random walk. 
Hubs in the ER graphs may emerge and reduce the saturation time in deterministic random walk, though these hubs are not as efficient as the hubs in the Barab\'{a}si-Albert (BA) model \cite{Barabsi1999} in stabilizing the saturation time.

This work has a broader impact beyond the random walk analysis as it highlights the importance of employing interactive environments for promoting innovative thinking. 
It has direct application where exploration and understanding of the interrelated yet immensely complex dynamics and underlying phenomena remain an existing challenge. 
For example, across several disciplines involving complex networks, such as electric power grids, telecommunication infrastructure, traffic networks, and interconnected vehicles.

\section{Methods}

\subsection{Deterministic Random Walk in NetLogo}

\begin{figure}[h!]
    \centering
    \includegraphics[scale=0.55]{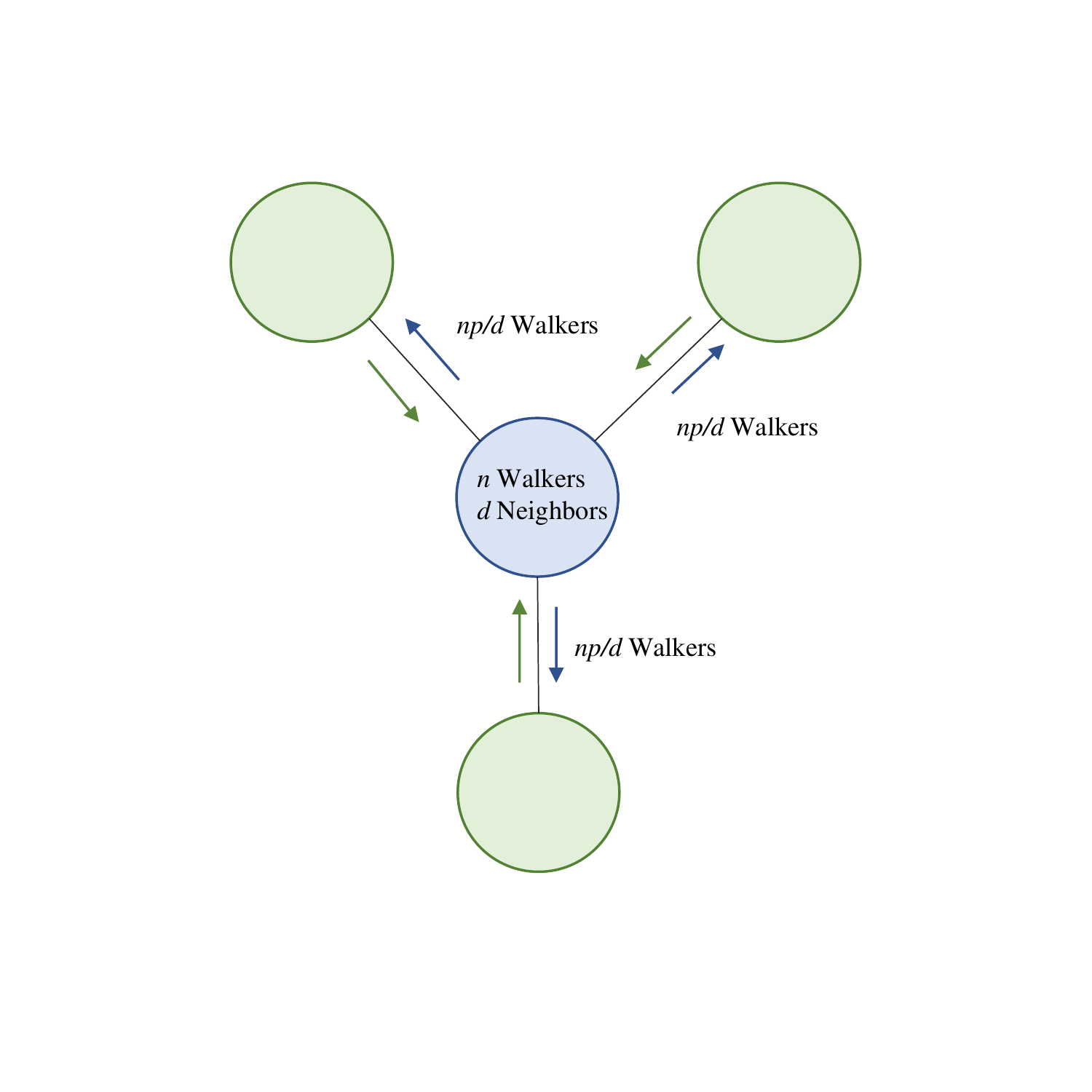}
    \caption{\textbf{Deterministic random Walk.} $p$ is the diffusion rate of the walkers. The blue node contains $n$ walkers and has $d$ neighbors at a certain instance of simulation. In this model, $n\times p$ walkers leave the blue node on an average and all its (green nodes) neighbors get $\frac{n\times p}{d}$ walkers from this node. In the meantime, some walkers are coming to the blue node from its (green nodes) neighbors using the same logic.}
    \label{fig:RW}
\end{figure}

The implementation of random walk processes on graph can be divided into two categories: deterministic and stochastic. 
The deterministic analysis allows the identification of baselines such as the stationary states in epidemic outbreaks \cite{barrat2008dynamical}, while the stochastic analysis provides us with diverse results that are generated from the same initial setting but are different due to the dynamic stochastic process. 
Under the same setting, even though the randomness of the stochastic random walk fluctuates the characteristics of dynamical processes like the distribution of the number of walks over nodes or the overall saturation time, the stochastic results will only fluctuate around the deterministic outcome. 
Here, we restrict our attention on the deterministic model of random walk. 
In the deterministic random walk model, partial walkers on a node leave the initial node, and all these leaving walkers equally distribute to all neighbors of this node (Fig. \ref{fig:RW}).

\begin{figure}[h!]
    \centering
    \includegraphics[width=\textwidth,height=\textheight,keepaspectratio]{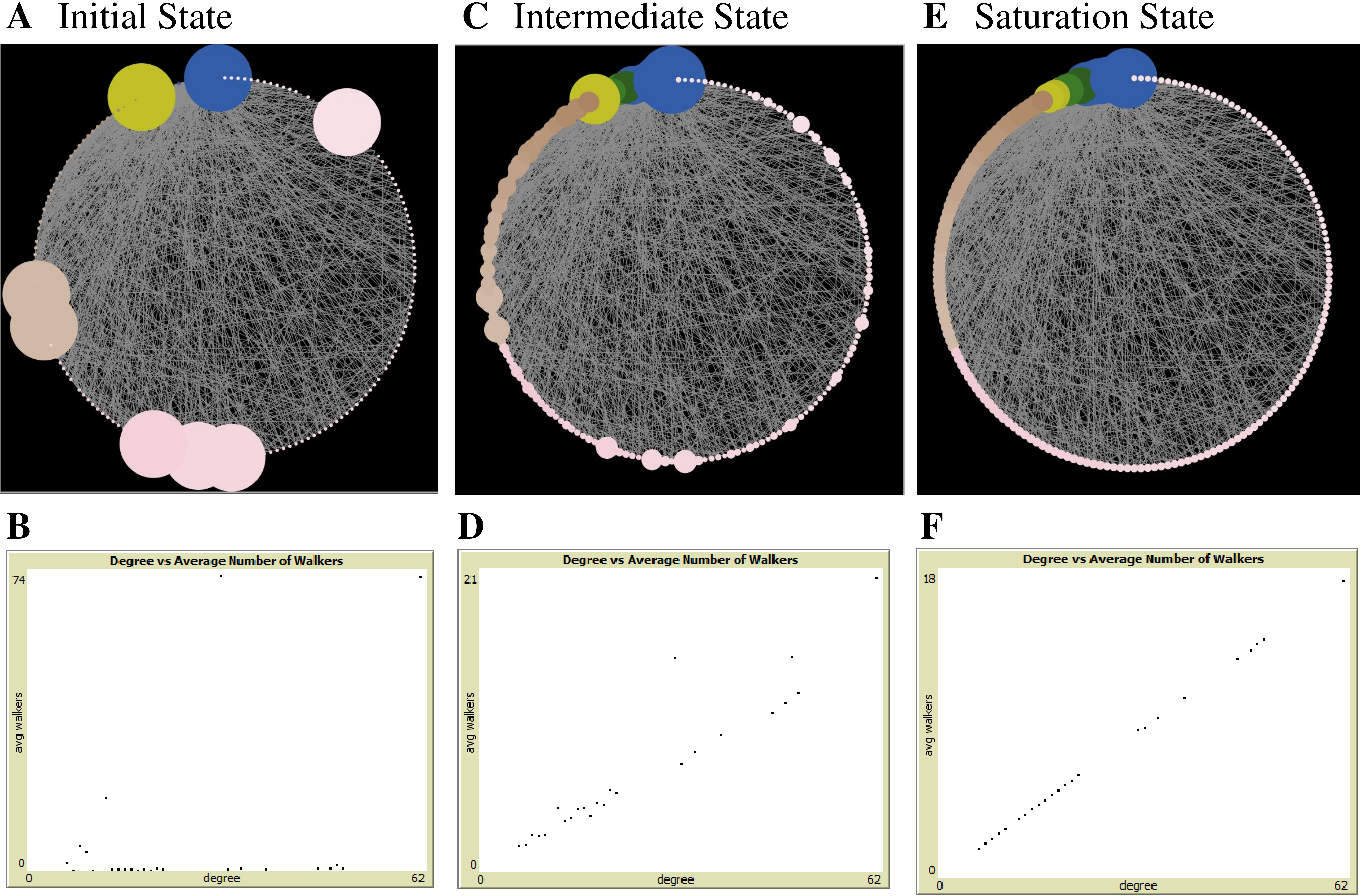}
    \caption{\textbf{Saturation Time in Random Walk.} \textbf{(A,C,E)} Circular layout in NetLogo frontend at the end of the random walk process on a BA graph with 180 nodes and a total of 600 walkers. Tunable parameters include type of network, number of nodes, average degree, layout type, total number of walkers, and initial number of nodes on which walkers are distributed randomly. \textbf{(B,D,F)} As the process saturates, the average walkers vs. degree plot slowly becomes a straight line.}
    \label{fig:fig_5_paper}
\end{figure}

The deterministic random walk over a network is realized after the following initialization: $m$ nodes are randomly picked and $n$ walkers are uniformly distributed in those $m$ nodes. 
After the initialization, walkers start to walk. 
In each time point $t$, walkers follow the rules in Fig.\ref{fig:RW} to move. 
Therefore, the number of walkers in node $i$ in time $t$ is, $n_i(t)$, updated by:

\begin{equation}
n_i(t) = n_i (t-1) \times (1-p)+\sum_{j}A_{ij} \times \frac{p \times n_j  (t-1)}{\sum_lA_{jl}}
\end{equation}
where $A$ is the adjacent matrix of the network and $p$ is the diffusion rate. As the random walk process progresses,  $n_i(t)$ constantly approaches to a value, that is:

\begin{equation}
 n_i(t) \to k_i \times \frac{n}{m}
\end{equation}
where $k_i$ is the degree of the $i$th node. 
Only at an equilibrium we observe a linear relation between the degree of nodes and the number of walkers, because $\frac{n}{m}$ is a constant (see Fig. \ref{fig:fig_5_paper}). 
Thus, for each time step, $t$, one can run a linear regression on the number of walkers of nodes to check whether or not the random walk process has arrived at an equilibrium. 
We set the necessary condition for arriving at an equilibrium to be $R^2(t) >0.99$, marking the saturation time when the processes reaches end.  
Thus, at the end of the realization, the number of the final walkers on a node depends only on its degree.

Next, we explain the implementation of these fundamental principles in NetLogo.

\subsection{Implementation in NetLogo}

Here we explain the steps and assumptions involving the back-end and front-end implementation of deterministic random walk model in NetLogo. 
Interactive front-end helps the users visualize the process in real-time, an important factor in teaching complex networks. 
Our back-end development is available to the public in the form of an open-source contribution to NetLogo. 
It presents various features convenient to users and suitable for both educational and research purposes, including the flexibility of tuning multiple parameters that influence the random walk process.

\subsubsection{Initializing the Network}

First step is to select the \emph{network-model}. 
This can be either ER or BA (scale-free) graphs. 
Next, the number of nodes in the network and the average degree is to select. 
Our model uses NetLogo algorithms to generate an instance of the input graph. 
The graph layout can be either a spring layout or a circular degree sorted layout. 
Coloring the nodes based on degree helps in an easy-to-follow visualization. 
We have defined two schemes for coloring the nodes. 
First, the \emph{degree single gradient} identifies the nodes from a lighter shade to a deeper one based on increasing degree. 
Second, the \emph{degree multi-bin gradient} scheme divides the nodes into different degree bins and assigns different colors to them. 
We posit this scheme is helpful to visualize the intermediate and the end states of the random walk process. 
As the random walk moves towards saturation, more walkers start moving towards the nodes with higher degrees. 
This means the higher degree nodes grow, hence, the size of the nodes are proportional to the number of walkers residing on them. 
Other tunable parameters include the diffusion probability (the rate at which a walker moves from one node to another), the total number of walkers, and the number of nodes on which the walkers are initially distributed. 
The degree distribution plot is either a Poisson distribution (for ER graphs) or a scale-free distribution (for BA graphs). 

\begin{figure}[b!]
    \centering
    \includegraphics[width=\textwidth,height=\textheight,keepaspectratio]{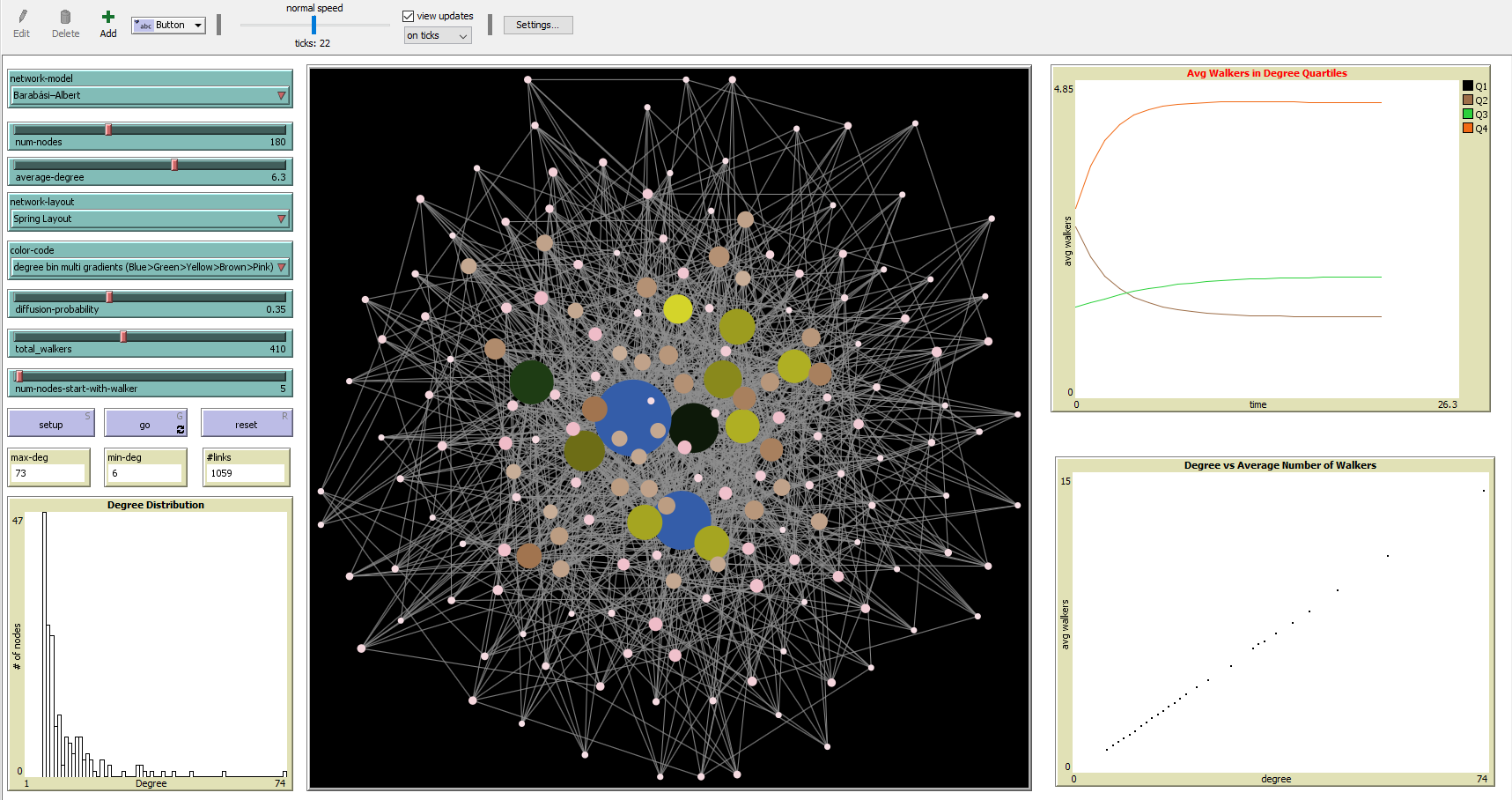}
    \caption{\textbf{NetLogo front-end for simulating random walk.} Left panel includes the tunable parameters. After setting these values, we run setup and the initial state of the process in created. While running the simulation, we observe the real-time behavior of the walkers both on the network and the plots on the right panel.
    The degree distribution is presented in the plot at the bottom left, the number of walkers in each degree quartile is visualized in the top right plot. 
    Lastly the relationship between the average number of walkers on nodes for each degree is shown in the bottom right plot.
}
    \label{fig:frontend}
\end{figure}

\subsubsection{Front-end implementation and interactive plots}

Our front-end implementation in NetLogo allows to visualize the network dynamics and the random walk processes in real-time. 
Fig. \ref{fig:frontend} shows a snapshot of our front-end implementation. 
As more walkers enter a node, the sizes of the nodes grow proportionally. 
We divide the nodes in four degree quartiles; therefore, the number of average walkers is displayed on each degree quartile. 
We observe that as the simulation progresses, the walkers move from the lower degree quartiles to the higher ones. 
Because of this reason, the average number of walkers and the degree shows a linear relationship, depicting as a straight line (Fig. \ref{fig:frontend} bottom-right plot). Our model computes the $R^2$ value of this straight line over time and when this value reaches $0.99$ threshold, it can be reasonably concluded that the process has saturated, i.e., the number of walkers on each node has reached an equilibrium.

\section{Results}

This section presents the results from the symmetry analysis of saturation time of the deterministic  in graphs, a problem that has not been studied nor reported in the literature to the best of our knowledge. 
Our analysis is carried out using our random walk implementation in NetLogo and aims to exhibit the scientific application of our development. 

We considered two case studies. 
First, we present the summary of our observations from NetLogo simulation on ER and BA graphs. 
In this case, we noticed an asymmetry in the saturation time of random walk on ER graphs. 
Accordingly, we establish there could exist instances in simulation where the process saturates sooner than the others in ER networks. 
In contrast, saturation time for BA graphs is monotonous. 
Second, we explore the frequency of this phenomenon in ER graphs, by we generating multiple ER graph instances and repeating the experiment. 
In this case, we observed that the occurrence of this asymmetric nature of saturation time is significantly frequent in ER graphs. 

\subsection{Asymmetric Saturation Time in ER Graphs}
\txr{This section analyzes the asymmetric saturation time of random walk in ER graphs. To this end, we generated an ER and a scale-free graph in NetLogo, running random walk simulations, and analyzing the observations. 
Both the ER and BA graphs had 1,000 nodes and an average degree of 6}. 
Then, we  uniformly distributed 400 walkers over 8 randomly selected nodes. 
We recorded the saturation time over 100 simulations and obtained a saturation time from each of the simulations. Using these 100 data-points, we computed the median saturation time, the $95^{th}$ percentile and the $5^{th}$ percentile values.  

\begin{figure}[]
    \centering
    \includegraphics[width=\textwidth,height=\textheight,keepaspectratio]{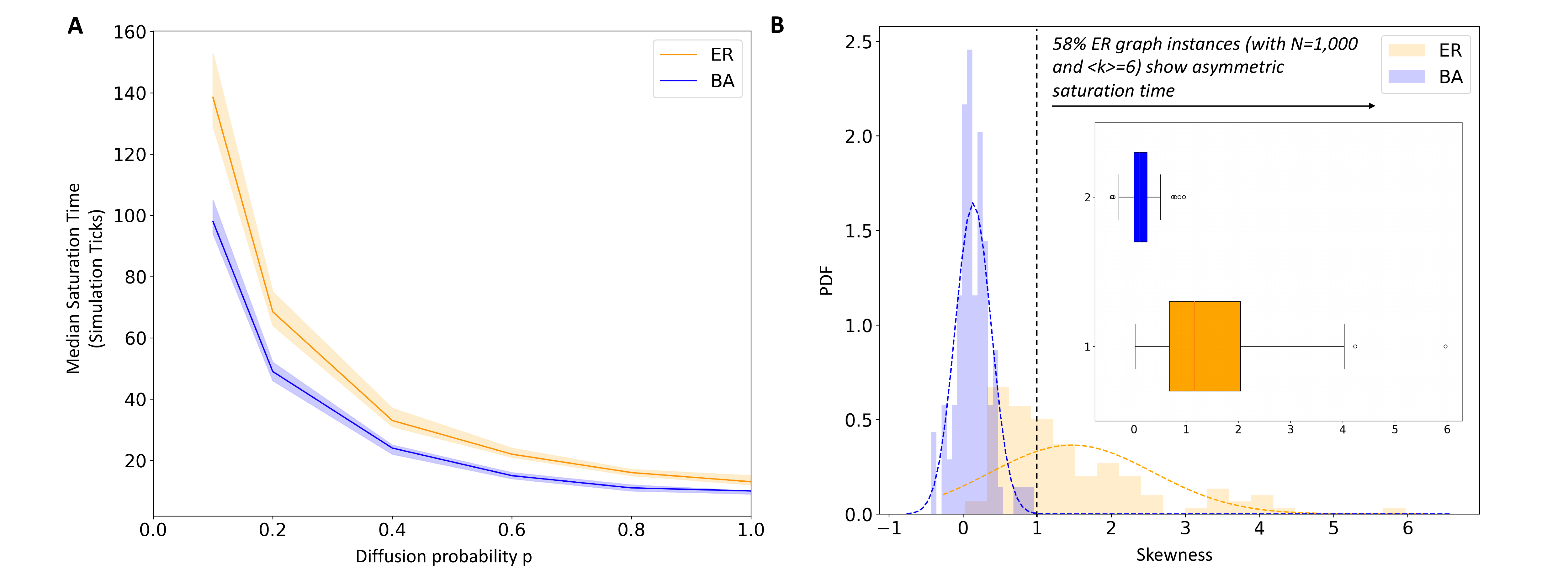}
    \caption{\textbf{Observing asymmetric saturation time in ER graphs.}  
    \txr{\textbf{(A)} Initially, we observed a wider upper bound of saturation time over 100 simulations in case of an ER and a BA graph (both with 1,000 nodes and 6 average degree). 
    This observation is consistent over different diffusion probabilities. 
    \textbf{(B)} To quantify the level of occurrence of this asymmetric behaviour, we studied the skewness of the distribution of saturation time for the ER and BA models. 
    Considering a high skewness threshold of 1.0, which is the upper-bound of skewness for BA graphs, we observed that the asymmetric saturation time occurred in 58\% of the ER graphs. 
    The distributions are generated through simulations over 100 instances of ER and BA graphs (each with 1,000 nodes and average degree of 6) by distributing 10,000 walkers evenly on 4 randomly selected nodes and repeating this process 100 times for each graph.}
    }
    \label{fig:results_combined}
\end{figure}

Throughout this case study, we observed a wider upper bound in case of ER graphs, as presented in Fig. \ref{fig:results_combined}A, whereas, in BA graphs, the saturation times are symmetrically distributed across the median value. 
We note that this observation is independent of the diffusion probably $p$ (Fig. \ref{fig:results_combined}A). Also, the asymmetric saturation time remained even in ER graphs at high $p$ values. This observation suggests a heterogeneous saturation time in the ER model such that some simulations over ER graphs take much lower time to saturate compared to others. Thus, we expect a wider right-tail of the saturation time distribution. 

Next, we systematically evaluated the consistency of this asymmetric behavior in ER graphs that we initially identified using the NetLogo environment.

\subsection{Consistency of Observing Asymmetric Saturation Time in ER Graphs}

\txr{This sections explores the generalizability of our observation on the emergence of ER graphs with asymmetric saturation time in deterministic random walks.
To this end, we generated 100 instances of ER and BA graphs. 
Each graph contained 1,000 nodes and had an average degree of 6. 
Then, we distributed 10,000 walkers randomly on 4 nodes and ran 100 such simulations on each of them to study the saturation time.}
For each distribution of the saturation time over a network, we calculated its skewness to measure the asymmetric symptom. 
Skewness is a measure of the asymmetry of the probability distribution of a real-valued random variable about its mean \cite{eberl2020asymptotic}. 
For a unimodal distribution, negative skew indicates that the tail is on the left side of the distribution, and positive skew indicates that the tail is on the right.

\txr{The results presented in Figure \ref{fig:results_combined}B indicate that the distribution of skewness in ER graphs has a longer positive tail. 
This positive skewness of the distribution suggests that on an average ER graphs could take longer to saturate than BA graphs. 
Considering a high skewness threshold of 1.0, which is the upper-bound of the skewness for BA graphs, we observed that the asymmetric saturation time \txr{occurred in 58\%} of the ER graphs.} 
Thus, our results suggest that the asymmetric nature of saturation time in random walk is plausibly related to the structural properties of the graphs. 

We will further explore the relationship between the saturation time and graph topology in the next section, Discussion.

\subsection{\txr{Network Topology and Asymmetric Saturation Time}}

\begin{figure}[h!]
    \centering    \includegraphics[clip,angle=0,width=1.0\textwidth]{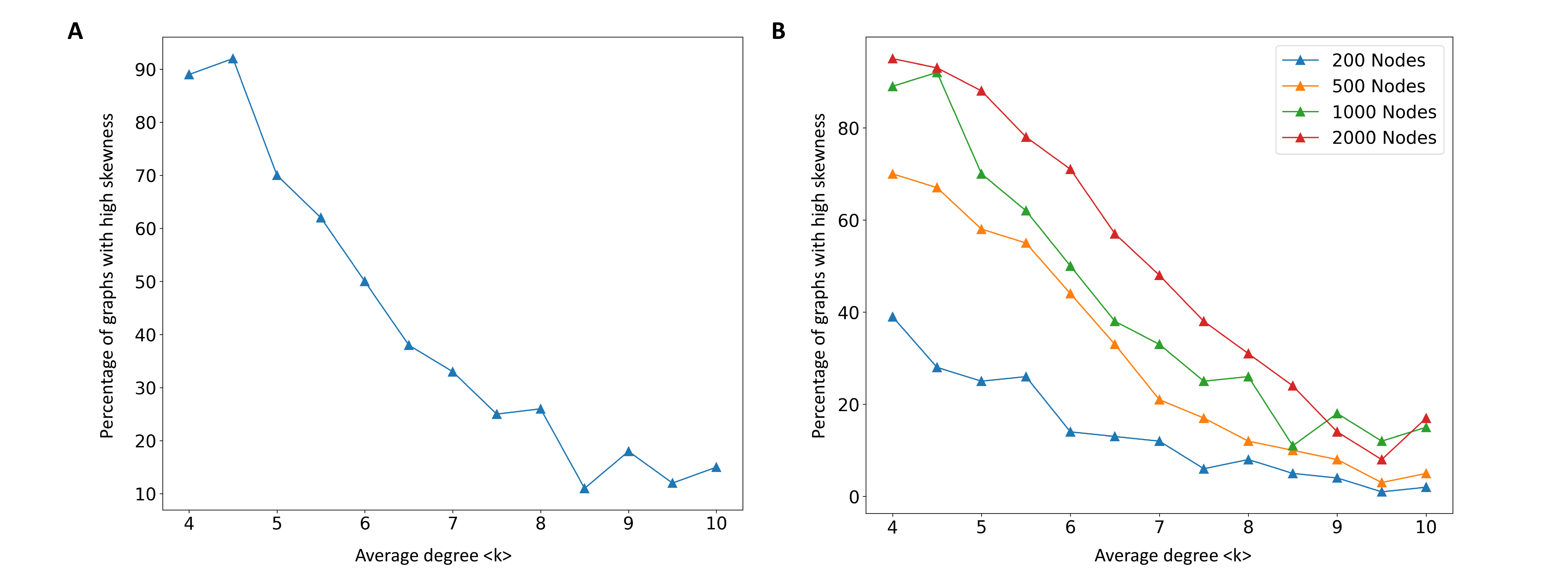}
    \caption{
\textbf{Consistency of Asymmetric Saturation Time in ER Graphs with Varying Size and Density.} 
\txr{
\textbf{(A)} 
We observe that as the graph becomes denser, less fraction of graph instances incur the asymmetric behavior in saturation time (saturation time distribution having skewness $\geq 1.0$). 
However, we consistently find instances of ER network with the observed asymmetric behaviour even in high average degree regime. 
For each average degree, We generated 100 ER graph instances with 1,000 nodes and ran 100 realizations of deterministic random walks to obtain the skewness of the saturation time. 
\textbf{(B)} Along with changing the average degree, we now vary the number of nodes. 
We observe consistent behavior regarding asymmetric saturation time across different numbers of nodes.}
    }
    \label{fig:shmoo_ER}
\end{figure}

This section explores the root cause of this asymmetric behaviour and uncover whether it is a byproduct of the network size or average degree. To this end, we conducted simulations with the same setting but varying $\langle k\rangle$ and $N$ in ER graphs. \txr{Based on the results shown in Fig. \ref{fig:shmoo_ER}, we conclude that the asymmetric saturation time is independent of the values of $\langle k\rangle$ and $p$, and our rationale is the following. The average shortest path distance or \emph{small-worldness} changes as $\Bar{d} \sim O (\frac{\log{n}}{\log \log{n}})$ in BA graphs, where $n$ is the number of nodes \cite{CHEN20081405}. On the other hand, the average shortest path distance grows faster in ER graphs with the number of nodes, following $\Bar{d} \sim O(\log n)$ \cite{Fronczak_2004}. Thus, with the increased number of nodes, ER graphs lose their small-worldness, and lack shortcuts between different compartments of a network. With reduced small-worldness, the random walkers take a larger time to saturate, bolstering the observations on asymmetric saturation time in Fig. \ref{fig:shmoo_ER}.}
%
%
%
We suggest that the observed asymmetric phenomenon has roots in the structure of ER graphs. 

Next, we will discuss this relationship to provide further insights and prospective.

\section{Discussion}

\begin{figure}[b!]
    \centering    \includegraphics[clip,angle=0,width=1.0\textwidth]{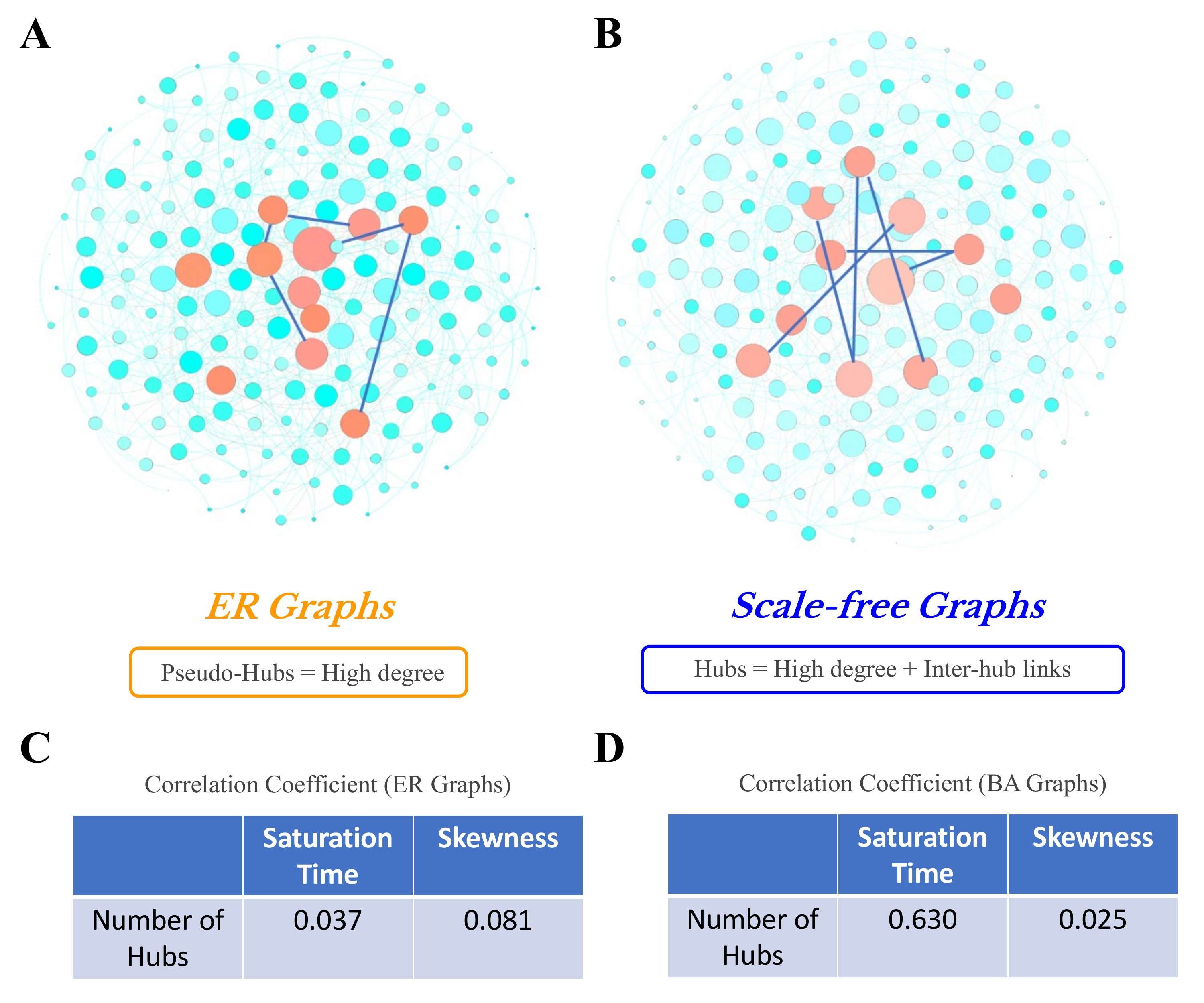}
    \caption{
    \textbf{Hubs and Asymmetric Saturation Time.} 
    \textbf{(A)} ER graphs have less inter-hub links, hence less shortcuts inside the network. 
    \textbf{(B)} Due to assortative mixing (hubs are linked), in BA graphs the hubs are more connected among themselves. 
    This creates many shortcuts for the walkers within different parts of the network that provide a stable saturation time for dynamical processes.
    \txr{
    \textbf{(C)} In ER graph the number of hubs (determined by the condition $deg(v)^2 > \sum_{j \in v.neighbors}{deg(j)}$) negatively correlate with the saturation time, meaning that having more hubs tend to decrease the saturation time. 
    But, having more hubs increases the instability as it positively correlates with the skewness of the saturation time, hence having asymmetric behavior. 
    \textbf{(D)} In contrast, in BA graphs we observe having more hubs is correlated with more stability in the saturation time (negative correlation with skewness), at the same time having more hubs increases the saturation time as it takes walkers more time to find nodes with very high degree. 
    The correlation coefficients in C-D are obtained by generating 100 different ER and BA networks with 1,000 nodes and 6 average degrees.
    Then, we ran 100 simulations (number of initial selected nodes is 4 and diffusion rate is 0.4) over each network thus, for each network, we found the skewness of the duration time distribution of these 100 simulations.
    }
    }
    \label{fig:Hubs_and_Pseudo_hubs}
\end{figure}



Thus far, we have demonstrated the educational and scientific application of our NetLogo implementation. 
We wish to further explore how this development could be leveraged to find answers to some of the open questions, extrapolate information, and draw inferences. 
To this end and further explore the possible underlying mechanisms that may cause the observed asymmetric phenomenon, we looked into the relationship between the degree distribution and the observed phenomenon, though the scientific capabilities of our development is not restricted only to the problem discussed here.

In BA graphs, the hubs are more connected among themselves due to assortative mixing. This creates many shortcuts among different parts of the network, decreasing the saturation time of dynamical processes \cite{pastor2015epidemic,keeling2005networks}. 
These additional inter-hub links in BA graphs enable walkers to find high-degree nodes robustly (Fig. \ref{fig:Hubs_and_Pseudo_hubs}A-B). This robust convergence of deterministic random walk can be explained by investigating the relationship between the number of hubs, median saturation time, and the observed asymmetric behaviour (skewness of saturation time distribution).
For demonstration purposes, we looked into these relationships by counting the number of hub-nodes in a network using the following condition: $deg(v)^2 > \sum_{j \in v.neighbors}{deg(j)}$, though we recognize a thorough analysis can be conducted as future work on these relationships.

\txr{Fig. \ref{fig:Hubs_and_Pseudo_hubs}C-D present the correlation between the number of hubs and the median and skewness of saturation time for BA and ER graphs. 
Surprisingly, we find that the Pearson's correlation coefficient between the number of hubs and median saturation time is 0.037 for ER graphs, but in BA it is 0.630, indicating that the BA hubs are more effective compared to ER in navigating the walkers to find high degree nodes.} 
In the BA model, having more hubs tends to increase the median saturation time, indicating that the walkers may need more time to find large-degree hubs to reach equilibrium (i.e., $R^2(t)>0.99$). 
In the case of ER graphs, the higher positive association between the number of hubs and skewness of saturation time signals that the structure of some ER graphs is less robust against outliers, being more susceptible to the initial selection of seeds that could trap walkers in the network's peripheral nodes and resulting the observed skewed distribution of saturation time. 

\section{Conclusions}

The interactive element and the temporal nature of NetLogo intrigued us to investigate the trajectory of saturation time in the deterministic random walk model in ER and BA models, leading us to find the asymmetric saturation behavior in ER graphs. 
Hence, we decided to promote network thinking in both teaching and research, we contributed the deterministic random walk model to NetLogo. 
This is the first model of dynamical processes in NetLogo over networks, and we envision that this work encourages more contributions related to dynamical processes to NetLogo in the future, like the stochastic random walk model \cite{barrat2008dynamical}. 

We believe that creating new toolkit with increased capabilities to model and contextualize complexity is essential for a successful development of the next generation of interdisciplinary thinkers, engineers, and scholars, a prerequisite for tackling the grand challenges of our century.

\section*{Acknowledgments}

We thank Professor A.~Vespignani and Dr. J.~Davis of Network Science Institute at Northeastern University for their insightful discussions and comments.

\section*{Declarations}

\subsection*{Ethical Approval}
not applicable.

\subsection*{Competing interests}
The authors have no competing interest.

\subsection*{Author contributions}

A.~C. contributed to the implementation of random walk in NetLogo, performing simulations in Python, creating figures and writing the manuscript.  
Q.~C. contributed to the implementation of random walk in NetLogo, performing simulations in Python, creating figures and writing the manuscript.  
A.~S. contributed to the problem statement and writing the manuscript. 
B.~R. contributed to designing of the problem statement, implementation of random walk in NetLogo, performing simulations, creating figures and writing the manuscript. 

\subsection*{Funding}
The authors have not received any funding for this project.

\subsection*{Code and Data Availability}

NetLogo and all the codes for reproducing the results presented in this paper are available open-source on Github via \hyperlink{https://github.com/bravandi/NetLogo-Dynamical-Processes}{https://github.com/bravandi/NetLogo-Dynamical-Processes}.

\bibliography{references}

\end{document}